\documentstyle[11pt]{article}

\title{\bf The pomeron-pomeron interaction\\ in the perturbative QCD}
\author{ N.Armesto and M.A. Braun\thanks{Visiting professor IBERDROLA.
Permanent address: Department of High-Energy Physics, University of St.
Petersburg, 198904 St. Petersburg, Russia.}
\\ Departamento de F\'{\i}sica de Part\'{\i}culas, 
\\ Universidade de Santiago de Compostela,\\ 15706-Santiago de
Compostela, Spain}
\date{June 1996}
\def\beq{\begin{equation}}
\def\eeq{\end{equation}}
\def\noi{\noindent}

\begin{document}
\maketitle
\medskip
\centerline{\bf Abstract}
\vskip 0.3cm

The pomeron-pomeron interaction is studied in the perturbative approach of
BFKL-Bartels. The total pomeron-pomeron cross-section is 
proportional to $\alpha_{s}^{4}s^{\Delta}/\sqrt {t_{1}t_{2}}$ where
$\sqrt{s}$ is the c.m. energy and $t_{1,2}$ are the virtualities of
the colliding pomerons. Upon calculating the coefficient the cross-section
is found to be of the order 2.2 mb at $\sqrt{s}= 6$ TeV and
$\sqrt{-t_{1}}\sim\sqrt{-t_{2}}\sim 1$ GeV/c.
\vspace{1cm}

\vspace*{7 cm}
\noi{\Large\bf US-FT/27-96}\\
{\Large\bf hep-ph/9606307}
\newpage
\section{Introduction}
The study of high-mass diffractive events brings into consideration
properties of the pomeron ($P$), which may be pragmatically associated
with a high-mass object  appearing between the rapidity gaps. In
particular, diffractive events in DIS allow to study the structure function
of the pomeron and central diffractive events in hadronic or photonic
collisions reveal the properties of the $PP$ interaction. Most of the
theoretical activity in this domain has  concentrated on the problem
of extracting the pomeron properties from the data, rather than to study
these properties themselves.

Such a study is possible in the perturbative framework of the "hard
pomeron", in which the realistic QCD is modified by introducing
an infrared cutoff (e.g. via the Higgs mechanism) and subsequently
fixing a (small) coupling constant. As is well known, this theory
results infrared stable in the colourless sector, so that one might think
that at least part of its predictions remain valid for the realistic QCD.

In a recent publication [1], in
this approach, we have studied the triple pomeron interaction,
which is vital for the pomeron structure function. The present note
extends this study to a more complicated case of the $PP$ scattering
cross-section, which evidently involves two triple pomeron vertices.

Since the hard pomeron theory is scale invariant, the dependence of the
$PP$ scattering cross-section on the relevant variables is trivially
predicted in this theory. If the two colliding pomerons have their momenta
$l$ and $l'$ with $t_{1}= l^{2}<0$ and 
$t_{2}= {l'}^{2}<0$, and their c.m. energy squared is
$s=(l+l')^{2}$, then from dimensional considerations and properties
of the BFKL pomeron [2] one finds for the total $PP$ cross-section
\beq
\sigma^{tot}_{PP}(s,t_{1},t_{2})=c\frac{\alpha_{s}^{4}s^\Delta}
{\sqrt{at_{1}t_{2}\ln s}}\ .
\eeq
Here the pomeron intercept $\Delta$ and the parameter $a$ are known
functions of the strong coupling constant $\alpha_{s}$ (see Eq. (18)) and
$c$ is an unknown number. Therefore the problem of studying the $PP$
cross-section reduces to finding the numerical constant $c$ in (1). One
should  have in mind that this constant factor is in fact rather
ill-defined. On the one hand, the separation of the $PP$ cross-section
from the rest of the amplitude inevitably involves a certain
arbitrariness, known as a choice of the "flux factor" [3]. In the hard
pomeron theory, as we shall see, this problem is aggravated because the
pomeron coupling to external sources results singular for small momentum
transfers. On the other hand, the scale of $\ln s$ in the hard pomeron
approach remains arbitrary, which introduces an undetermined factor in (1).

In the following we calculate the coefficient $c$ within a certain
(rather obvious) choice of the flux factor and neglecting the numerical
factor which comes from $s^{\Delta}$ in (1). This latter approximation
corresponds to the (theoretically justified) assumption of a very small
$\Delta$. In applications we take the scale involved of the order of 1
GeV$^{2}$, as favoured by the pomeron phenomenology. Our calculations
give a rather large value for $c$:
\beq
c\simeq 35370.
\eeq

Possible experimental consequences of this result are discussed in
Section 4 of this note. Sections 2 and 3 are dedicated to its derivation.

\section{The central diffraction cross-section}

The $PP$ cross-section enters as a part of a more general central
diffraction cross-section in which the colliding hadrons (or photons)
produce a high mass $M$ in the central region separated from the target
and projectile by large rapidity gaps (Fig. 1). As seen from this figure, both
projectile and target emit pomerons with momenta $l$ and $l'$,
respectively, which interact in the central part of the diagram. We
introduce the standard energetic variables (see Fig. 1 for the notation of
the momenta involved)
\beq
s=(p_{1}+p_{2})^{2},\ \ s_{12}=(p_{1}+l')^{2},\ \ 
s_{21}=(p_{2}+l)^{2},\ \ M^{2}=(l+l')^{2},
\eeq
constrained by the relation
\beq
s_{12}s_{21}=sM^{2}.
\eeq
We assume that all these variables are large with
$s>>s_{12},s_{21}>>M^{2}$. We also introduce
\beq
s_{1}=s_{21}/M^{2},\ \ s_{2}=s_{12}/M^{2},
\eeq
which serve as energetic variables for the upper and lower pomerons in
Fig. 1, respectively.

In the hard pomeron approach the pomerons which connect the central part
of Fig. 1 with the target and projectile are identified as  BFKL
pomerons. As to the central part itself, it is also given by a
BFKL pomeron coupled to the upper and lower pomerons by two Bartels
vertices $K_{2\rightarrow 4}$, which describe transition of two
reggeized gluons into four [4] (see Fig. 2). Explicitly the upper vertex
 $K_{2\rightarrow 4}$, apart from a
colour factor and a factor $g^{4}$, is given by
\[
K_{2\rightarrow 4}(q_{1},-q_{1};q_{2},l-q_{2},-l+q_{3},-q_{3})\equiv
K_{l}(q_{1},q_{2},q_{3})=\]\beq
\frac{q_{1}^{2}q_{2}^{2}}{(q_{1}-q_{2})^{2}}+
\frac{q_{1}^{2}q_{3}^{2}}{(q_{1}-q_{3})^{2}}-
\frac{q_{1}^{4}(q_{2}-q_{3})^{2}}{(q_{1}-q_{2})^{2}(q_{1}-q_{3})^{2}}\ .
\eeq
The lower vertex has the same form with the primed momenta.

Then we obtain the expression for the absorptive part $D$ corresponding to
Fig. 1 in the form 
\[
D=(1/16)g^{16}N^{4}(N^{2}-1)(s^{2}/M^{2})(2\pi)^{-12}
\int \prod_{i=1}^{3}(d^{2}q_{i}d^{2}q'_{i})K_{l}(q_{1},q_{2},q_{3})
K_{l'}(q'_{1},q'_{2},q'_{3})\]
\beq
\phi_{1}(s_{1},l,q_{2})\phi_{1}(s_{1},-l,-q_{3})
\phi_{2}(s_{2},l',q'_{2})\phi_{2}(s_{2},-l',-q'_{3})
G(M^{2},0,q_{1},-q'_{1}).
\eeq
Here $\phi_{1(2)}(s,l,q)$ is a pomeron (a solution to the BFKL equation)
 coupled
to the projectile (target) colourless external source with energy
squared  $s$,
total momentum $l$ and one of the gluon's momentum $q$.
The function $G(M^{2},0,q_{1},q'_{1})$ is the BFKL Green function for the 
energy squared $M^{2}$, total momentum zero and initial and final
momentum of one of the gluons $q_{1}$ and $q'_{1}$, respectively.
It is assumed
that the colour factor for each source is $(1/2)\delta_{ab}$ and that
 each source
is proportional to $g^{2}$. The factor $(1/16)N^{4}(N^{2}-1)$
comes from the colour variables, $N$ being the number of colours ($N=3$
for the physical QCD). Our normalization is
that the double inclusive cross-section described by Fig. 1 is given by
\beq \frac{d\sigma}{dt_{1}dt_{2}ds_{12}ds_{21}}=\frac{D}{256\pi^{4}s^{3}}\ .
\eeq

Since for $l\neq 0$ the solution of the BFKL equation is easier to obtain
 in the (transversal) coordinate space, we pass to this space by
presenting
\beq
\phi(s,l,q)=\int d^{2}r\phi(s,l,r)\exp [ir(q-l/2)].
\eeq
Evidently $r$ is the transversal distance between the gluons. Then (7)
transforms into
\[
D=(1/16)g^{16}N^{4}(N^{2}-1)(s^{2}/M^{2})
\int \prod_{i=1}^{3}(d^{2}r_{i}d^{2}r'_{i})K_{l}(r_{1},r_{2},r_{3})
K_{l'}(r'_{1},r'_{2},r'_{3})\]
\[\exp [-il(r_{2}+r_{3})/2-il'(r'_{2}+r'_{3})/2]\]\beq
\phi_{1}(s_{1},l,r_{2})\phi_{1}(s_{1},-l,-r_{3})
\phi_{2}(s_{2},l',r'_{2})\phi_{2}(s_{2},-l',-r'_{3})
G(M^{2},0,r_{1},-r'_{1}),
\eeq
where the 2 to 4 reggeon vertex in the coordinate space is obtained from
(6) to be
\[
K_{l}(r_{1},r_{2},r_{3})=
-2\left[\delta^{2}(r_{2})
\delta^{2}(r_{3})\nabla_{1}^{2}\delta^{2}({r})+
\frac{1}{2\pi}\delta^{2}(r_{3})r_{2}^{-2}(r_{2}\nabla_{1})
\nabla_{1}^{2}\delta^{2}({r})+\right.\]\beq\left.
\frac{1}{2\pi}\delta^{2}(r_{2})r_{3}^{-2}(r_{3}\nabla_{1})
\nabla_{1}^{2}\delta^{2}({r})+\frac{1}{(2\pi)^{2}}
r_{2}^{-2}r_{3}^{-2}(r_{2}r_{3})\nabla_{1}^{4}\delta^{2}({r})\right],
\eeq with ${r}=r_{1}+r_{2}+r_{3}$. The solutions $\phi$ vanish when
$r=0$. So only the last term in (11) survives.
We then can rewrite (10) as
\[
D=(1/4)g^{16}N^{4}(N^{2}-1)(s^{2}/M^{2})(2\pi)^{-8}\]\beq
\int d^{2}qd^{2}q'\chi(M^{2},q+l/2,q'+l'/2)
[\nabla_{q}\chi_{1}(s_{1},l,q)]^{2}
[\nabla_{q'}\chi_{2}(s_{2},l',q')]^{2},
\eeq
where
\beq
\chi(s,q,q')=\int d^{2}rd^{2}r'\exp (iqr+iq'r')\nabla^{4}{\nabla'}^{4}
G(M^{2},0,r,-r')
\eeq
and
\beq
\chi_{1,2}(s,l,q)=\int d^{2}r r^{-2}\phi_{1,2}(s,l,r)\exp(iqr).
\eeq

To calculate the function $\chi(s,q,q')$ we first integrate by parts
in (13) to remove the derivatives,
\beq
\chi(s,q,q')=(qq')^{4}\int d^{2}rd^{2}r'
G(M^{2},0,r,-r')\exp (iqr+iq'r').
\eeq
The BFKL Green function  at
$l=0$ and large $s$ is given by the expression [5]
\beq
G(s,0,r,-r')=\frac{1}{32\pi^{2}}rr'\int_{-\infty}^{\infty}\frac{d\nu
s^{\omega(\nu)}} {(\nu^{2}+1/4)^{2}}(r/r')^{-2i\nu},
\eeq
where
\beq
\omega(\nu)=(g^{2}N/2\pi^{2})[\psi(1)-{\rm Re}\psi(1/2+i\nu)].
\eeq
Small values of $\nu$ play the dominant role in (16) at large $s$, so
that we can approximate
\beq
\omega(\nu)=\Delta-a\nu^{2};\ \ \Delta=(g^{2}N/\pi^{2})\ln 2,\ \
a=(7g^{2}N/2\pi^{2})\zeta (3).
\eeq
Calculating the integrals over $r$ and $r'$ in (15) and then the remaining
integral over $\nu$ we obtain 
\beq
\chi(s,q,q')=2qq's^{\Delta}\sqrt{\frac{\pi}{a\ln s}}\exp\left(
-\frac{\ln^{2} (q/q')}{a\ln s}\right).
\eeq
At large $s$ the exponential factor in (19)  can evidently be
neglected.

Now we turn to the functions $\chi_{1,2}(s,l,q)$.
The solutions $\phi_{1(2)}$ can be obtained by using the Green function
of the BFKL equation for a given total momentum $G(s,l,r,r')$. 
Following [6] and taking into account that the Green function $G$
vanishes if $r$ or $r'$ are equal to zero and is isotropic at high
energies one finds for, say,  the projectile: \beq
\phi_{1}(s,l,r)=-2\int d^{2}\tilde{r}G(s,l,r,\tilde{r})
\rho_{1}(l,\tilde{r})\exp (il\tilde{r}/2).
\eeq
Here $\rho_{1}(l,r)$ is the colour density of the projectile as a function
of the intergluon distance $r$ with the colour factor $(1/2)\delta_{ab}$
and $g^{2}$ separated. The explicit form of $\rho$ can easily be found if
the projectile is a highly virtual photon. For the longitudinal photon
then
\beq
\rho^{(L)}_{1}(r)=\frac{4e^{2}|p_{1}^{2}|}{(2\pi)^{3}}
\sum_{f=1}^{N_{f}}Z_{f}^{2}\int_{0}^{1} d\alpha[\alpha
(1-\alpha)]^{2}{\rm K}_{0}^{2}(\epsilon_{f} r)\exp (-i\alpha lr),
\eeq
where $\epsilon_{f}^{2}=|p_{1}^{2}|\alpha (1-\alpha)+m_{f}^{2}$ and $m_{f}$
and $Z_{f}$ are the mass and charge of the quark of flavour $f$, respectively.
For the transverse photon a slightly more complicated formula emerges
\[
\rho^{(T)}_{1}(r)=\frac{e^{2}}{(2\pi)^{3}}
\sum_{f=1}^{N_{f}}Z_{f}^{2}\int_{0}^{1}d\alpha\]\beq
\left\{m_{f}^{2}{\rm
K}_{0}^{2}(\epsilon_{f} r)
+[\alpha^{2}+(1-\alpha)^{2}]\epsilon_{f}^{2}{\rm K}_{1}^{2} (\epsilon_{f}
r)-\alpha (1-\alpha)(1-2\alpha)
{\rm K}_{0}(\epsilon_{f}r)il\nabla_{r}{\rm K}_{0}(\epsilon_{f}r)\right\}
\exp(-i\alpha lr). \eeq
Eqs (21), (22) generalize the well-known formulas
for zero momentum transfer [6]. For a hadron projectile or target the form
of the density is, of course, unknown, although one expects that it should
be less singular at $r=0$ and normalizable.
 
The leading contribution
to the BFKL Green function at $l\neq 0$ has the form [5]
\beq
G_{l}(s,r,r')=\frac{1}{(2\pi)^{4}}\int\frac{d\nu
\nu^{2}}{(\nu^{2}+1/4)^{2}} s^{\omega(\nu)}E_{l}^{\nu}(r)E_{l}^{-\nu}(r'),
\eeq
where
\beq
E_{l}^{\nu}(r)=\int
d^{2}R\exp(ilR)\left(\frac{r}{|R+r/2||R-r/2|}\right)^{1+2i\nu}. \eeq
It is evident that the Green function (23), transformed into momentum
space, contains terms proportional to $\delta^{2}(l/2\pm q)$, which
should be absent in the physical solution (this circumstance was first
noted by A.H.Mueller and W.-K.Tang [7]). For that, (23) goes to zero
at $r=0$. Terms proportional to $\delta^{2}(l/2+q)$ are not dangerous
to us: they are killed by the vertex $K_{2\rightarrow 4}$.
 To remove the dangerous singularity at $q=l/2$
and simultaneously preserve good behaviour at $r=0$ we therefore make
a subtraction in $E$, changing it to
\beq
\tilde{E}^{\nu}_{l}(r)=
\int d^{2}R\exp(ilR)\left[\left(\frac{r}{|R+r/2||R-r/2|}\right)^{1+2i\nu}-
|R+r/2|^{-1-2i\nu}+|R-r/2|^{-1-2i\nu}\right].
\eeq
This subtraction removes the $\delta$ singularity at $q=l/2$ and doubles
the one at $q=-l/2$, the latter eliminated by the kernel
$K_{2\rightarrow 4}$.

Integration over $r$ in (14) leads to the integral
\beq
J(l,q)=\int (d^{2}r/(2\pi)^{2})(1/r^{2})\tilde{E}^{\nu}_{l}(r)\exp(iqr).
\eeq
This integral is convergent at any values of $\nu$, the point $\nu=0$
included, when the convergence at large values of $r$ and $R$ is
provided by the exponential factors. So in the limit
$s\rightarrow\infty$, when small values of $\nu$ dominate, we can put
$\nu=0$ in $J$:
\beq
J(l,q)=\int(d^{2}Rd^{2}r/(2\pi)^{2})\frac{\exp(ilR+iqr)}
{r|R+r/2||R-r/2|}+(1/l)\ln\left(\frac{|l/2-q|}{|l/2+q|}\right).
\eeq
The second term comes from the subtraction terms in (25). Passing to the
Fourier transform of the function $1/r$ one can represent (27) as an
integral in the momentum space
\beq
J(l,q)=(1/l)\int \frac{d^{2}p
(l+|l/2+p|-|l/2-p|)}{(2\pi)|l/2+p||l/2-p||q+p|}\ . \eeq

In the same manner we can put $\nu=0$ in the function
$E^{-\nu}_{l}(\tilde{r})$ obtaining for the integral over the transverse
dimensions of the projectile \beq
\int
\frac{d^{2}Rd^{2}r\exp[il(R+r/2)]r\rho_{1}(l,r)}{|R+r/2||R-r/2|}=
\pi R_{1}F_{1}(t_{1}). \eeq
Here we have separated the characteristic transverse dimension of the
projectile
\beq
R_{1}=\int d^{2}r\,r\rho_{1}(0,r)
 \eeq
(and a factor $\pi$ for convenience) and introduced a
dimensionless function $F_{1}(t_{1})$, which is a vertex for
the interaction of the projectile with a pomeron at momentum transfer
$\sqrt{-t_{1}}=l$. $F_{1}(t_{1})$ behaves like $\ln |t_{1}|$ at small
$t_{1}$. The rest of the Green function is easily calculated by the
stationary point method to finally give \beq
\chi_{1}(s,l,q)=-\frac{4s^{\Delta}}{\sqrt{\pi}(a\ln
s)^{3/2}}R_{1}F_{1}(t_{1})J(l,q). \eeq

Evidently the function $\chi_{2}(s,l,q)$ is given by the same formula
with the projectile form-factor $F_{1}$ substituted by the target one
$F_{2}$.

Combining our results for $\chi$ and $\chi_{1,2}$,  we obtain
our final expression for the absorptive part $D$ corresponding to the
central diffraction from Eq. (12) to be\[
D=8\pi^{-11/2}g^{16}N^{4}(N^{2}-1)(s^{2}/M^{2})
\frac{(s_{1}^{2}s_{2}^{2}M^{2})^{\Delta}}{\sqrt{a\ln M^{2}}
(a\ln s_{1})^{3}(a\ln s_{2})^{3}} \]\beq
[R_{1}R_{2}F_{1}(t_{1})F_{2}(t_{2})]^{2}B^{2}/\sqrt{t_{1}t_{2}}. \eeq
Here the number $B$ is defined as a result of the $q$ integration
\beq
B=l\int (d^{2}q/(2\pi)^{2})|l/2+q|[\nabla_{q}J(l,q)]^{2}.
\eeq
It does not depend on $l$ and can be represented as an integral over
three momenta
\beq
B=(1/(2\pi )^{4}l)\int d^{2}qd^{2}pd^{2}p'
\frac{|l/2+q|(q+p)(q+p')(l+p_{+}-p_{-})(l+p'_{+}-p'_{-})}
{p_{+}p_{-}p'_{+}p'_{-}|q+p|^{3}|q+p'|^{3}}\ ,
\eeq
where
 \beq p_{\pm}=|p\pm l/2|,\ \ p'_{\pm}=|p'\pm l/2|.\eeq
It is a well-defined integral. Calculations give
\[ B=1.962\pm 0.002.\]

The remaining problem is to separate the $PP$ cross-section from the
absorptive part $D$.

\section{The pomeron-pomeron cross-section}
To separate the $PP$ cross-section from the absorptive part for central
diffractive events we compare the latter with the elastic amplitude $A(s,t)$
for the scattering of the projectile and target in the same approach.
In the single pomeron exchange
approximation we have
\beq
A(s,t)=ig^{4}(N^{2}-1)s\int d^{2}rd^{2}r'\rho_{1}(l,r)
\rho_{2}(-l,r')G_{l}(s,r,r').
\eeq
The normalization is chosen to have the elastic cross-section 
\beq
d\sigma^{el}/dt=(1/16\pi s^{2})|A(s,t)|^{2}.
\eeq

Performing calculations similar to which lead us to Eq. (32) we find 
for the forward scattering $t=0$
\beq
A(s,0)=i\frac{2}{(2\pi)^{2}}g^{4}(N^{2}-1)s^{1+\Delta}
R_{1}R_{2}\sqrt{\frac{\pi}{a\ln s}}
\eeq
and for $t<0$
\beq
A(s,t)=i\frac{2}{(2\pi)^{2}}g^{4}(N^{2}-1)s^{1+\Delta}R_{1}R_{2}
F_{1}(t)F_{2}(t)
\frac{\sqrt{\pi}}{(a\ln s)^{3/2}}\ .
\eeq
Evidently one cannot determine the pomeron and its couplings in a unique
way from these expressions. One also notices that the  
amplitude at $t=0$ rises with $s$ faster that at $t<0$ by a factor $\ln
s$. This shows that the amplitude has a discontinuous behaviour at $t=0$,
which results in the logarithmic singularity of the form-factors
$F_{1,2}(t)$ at $t=0$. As a consequence we meet with an additional
difficulty in fixing the "flux factor", since working at $t<0$, we cannot
fix the coupling by its value at $t=0$.

Therefore rather to fix the flux factor from the coupling, we choose a
particular expression for the pomeron propagator, which simplifies the
factor entering the triple pomeron vertex. We take that the pomeron
contribution is at $t=0$ \beq
P(s,0)=2\sqrt{\pi}\frac{s^{1+\Delta}}{\sqrt{a\ln s}}\eeq and at $t<0$
\beq P(s,t)=4\sqrt{\pi}\frac{s^{1+\Delta}}{(a\ln s)^{3/2}}\ .\eeq
 Then we find
for its coupling $\gamma$ to a real external particle at $t=0$
\beq
\gamma(0)=\frac{1}{2\pi}g^{2}\sqrt{N^{2}-1}\ R
\eeq
and at $t<0$
\beq
\gamma(t)=\frac{1}{2\pi\sqrt{2}}g^{2}\sqrt{N^{2}-1}\ RF(t).
\eeq
These expression define the flux factor in our case. Of course, they are
not unique: one can always multiply the coupling by a constant,
simultaneously dividing the propagator by its square.

Rewriting (32) in terms of these quantities we obtain
\beq
D=M^{2}\gamma_{1}^{2}(t_{1})\gamma_{2}^{2}(t_{2})
P^{2}(s_{1},t_{1})P^{2}(s_{2},t_{2})\sigma^{tot}_{PP}(M^{2},t_{1},t_{2}),
\eeq
where the $PP$ total cross-section,  defined by this equation, is
\beq
\sigma^{tot}_{PP}(M^{2},t_{1},t_{2})=
2\pi^{-7/2}g^{8}\frac{N^{4}}{N^{2}-1}\frac{M^{2\Delta}}{\sqrt{a\ln
M^{2}}}\frac{B^{2}}{\sqrt{t_{1}t_{2}}}\ . 
\eeq
Evidently it has the form (1) indicated in the Introduction on general
grounds, with the constant  factor $c$ given by 
\beq
c=512\sqrt{\pi}\frac{N^{4}}{N^{2}-1}B^{2}\simeq 35370.
\eeq

Note that the $PP$ cross-section can be written in terms of the
triple-pomeron vertex $\gamma_{3P}(t)$ found in [1] to be (with the
flux factor given by Eqs. (40)-(43))
\beq
\gamma_{3P}(t)=\frac{g^{4}N^{2}B}{\pi^{2}\sqrt{N^{2}-1}}
\frac{1}{\sqrt{-t}}\ .
\eeq
In terms of $\gamma_{3P}$ and the pomeron propagator at $t=0$ (40) the $PP$
cross-section (45) reads
\beq
\sigma^{tot}_{PP}(M^{2},t_{1},t_{2})=\gamma_{3P}(t_{1})\gamma_{3P}(t_{2})
P(M^{2},0)/M^{2}.
\eeq
Thus the pomeron-pomeron interaction
correctly factorizes  into a pair of triple pomeron vertices
joined by a pomeron which propagates between them (this fact is
independent of the normalization of the pomeron propagators in Eqs. (40)
and (41)). The characteristic $1/\sqrt{t_{1}t_{2}}$ behaviour of the
cross-section follows that of the two triple pomeron vertices.

\section{Discussion}

The obtained result corresponds to an exact prediction of the hard
pomeron theory, within the mentioned uncertainties associated with the
choice of scale in logarithmic factors. To relate these results to
observable data one has to specify the range of transferred momenta to
which they may apply and the value of the fixed coupling constant
$\alpha_{s}$. The transferred momenta should not evidently be too low,
to avoid confinement effects neglected in the hard pomeron theory. So 
one should take at least $|t_{1,2}|>1$ (GeV/c)$^{2}$ and higher values of 
$|t_{1,2}|$ would be preferable. 

As to the coupling constant $\alpha_{s}$, it may naively be identified
with the expected value of the running QCD coupling constant at the
appropriate scale defined by the values of $|t_{1,2}|$. However assuming
that $\alpha_{QCD}$ is around 0.2 at 1 GeV/c, one arrives at the
well-known value of the intercept $\Delta\sim 0.5$, which  not only 
 contradicts soft scattering data but  is also about twice larger that
the intercept observed in low -$x$ DIS data, performed at comparable
values of the momentum transfer  [8]. For this reason we prefer to treat
$\alpha_{s}$ as a free parameter inherent in the hard pomeron theory,
whose values can be deduced from the observed intercepts determined at
appropriate momentum transfers. 
Low-$x$ DIS experiments at relatively small momentum transfers give the
value of the intercept $\Delta\simeq 0.19$ [8]. From this, using
(18) we deduce
\beq
\alpha_{s}=0.072.
\eeq

With this value of $\alpha_{s}$ and at $|t_{1,2}|= 1$ (GeV/c)$^{2}$ we get
for the triple pomeron vertex (Eq. (47))
\beq
\gamma_{3P}(|t|=1\ ({\rm GeV/c})^{2})=0.32\ {\rm mb}^{1/2}
\eeq
and for the $PP$ total cross-section
\beq
\sigma^{tot}_{PP}(s,|t_{1,2}|= 1\
({\rm GeV/c})^{2})=0.34\frac{s^{0.19}}{\sqrt{\ln s}}\ {\rm mb}.
\eeq
In the latter formula we assume that the appropriate scale is around 1
GeV, so that $s$ should be measured in GeV$^{2}$. At $\sqrt{s}=6$ TeV 
from (51) we get the $PP$ cross-section 2.2 mb.

Comparing to the existing phenomenological estimates, based on
diffractive scattering data, we observe that the value of the triple
pomeron vertex (50) is quite close to the one obtained from the
lower energy and $t$ data  $\gamma_{3P}=0.364\pm 0.025$ mb$^{1/2}$ [9].
The experimental value actually refers to quite small $|t|\sim 0.05$
(GeV/c)$^{2}$. Our theoretical value obtained for much higher $|t|$ is 
only slightly
smaller, which seem to indicate that the triple pomeron interaction
freezes at $|t|$ of the order of $1$ (GeV/c)$^{2}$, where the confinement
effects are expected to set in.

As a result,  our predictions of the $PP$ cross-section are
of the same order as the estimates done in [10] on the basis of the
experimental value for the triple pomeron vertex. In this reference values
of $\sigma_{PP}$ for the $PP$ c.m. energy intervals $2\div 180$ and $2\div
600$ GeV were estimated to be 0.3 mb and 0.6 mb, respectively.
Our Eq. (51) gives $\sigma_{PP}=0.37$ mb for $\sqrt{s}=2$
GeV and 1.08 mb for $\sqrt{s}=600$ GeV.

One
should have in mind that the magnitude of the $PP$ cross-section is very
sensitive to the choice of $\alpha_{s}$ and the resulting values of the
intercept $\Delta$. Taking $\alpha=0.1$, only slightly larger than (49),
raises the intercept to $\Delta=0.265$,  with a consequence that the $PP$
cross-section  at $\sqrt{s}=6$ TeV and $|t_{1,2}|= 1$ (GeV/c)$^{2}$ rises
more than 10 times to achieve values around 30 mb! For the range of
energies considered in [10] we would then obtain PP cross-sections
between 1.3 mb (at 2 GeV) and 9 mb (at 600 GeV). Therefore
the experimental study of the $PP$ cross-section presents a very efficient
and sensitive test for possible signatures of the hard pomeron physics.

\section{Acknowledgements}
The authors express their deep gratitude to Prof. C.Pajares for
illuminating discussions. N.A. and M.A.B. thank the CICYT of Spain and
IBERDROLA, respectively, for finantial support. M.A.B. is also greatful to
INTAS, which partly supported this study under the project INTAS 93-79.

\newpage \section{References}

\noi [1] M.A.Braun, St. Petersburg University preprint
SPbU-IP-1995/10 (hep-ph/9506245) (to be published in Z. Phys. {\bf C}).

\noi [2] V.S.Fadin,
E.A.Kuraev and L.N.Lipatov, Phys. Lett. {\bf B60} (1975) 50;
I.I.Balitsky and L.N.Lipatov, Sov. J. Nucl. Phys. {\bf 15} (1978) 438.

\noi [3] A.Donnachie and P.V.Landshoff, Nucl. Phys. {\bf B244} (1984) 322;
E.L.Berger, J.C.Collins, D.E.Soper and G.Sterman, Nucl. Phys. {\bf B286}
(1987) 704;
K.Goulianos, Phys. Rep. {\bf 101} (1983) 169.

\noi [4] J.Bartels, Nucl. Phys. {\bf B175} (1980) 365.

\noi [5] L.N.Lipatov, Zh. Eksp. Teor. Fiz. {\bf 90} (1986) 1536 (Sov. Phys.
JETP {\bf 63} (1986) 904).

\noi [6] N.N.Nikolaev and B.G.Zakharov, Z. Phys. {\bf C49} (1991) 607.

\noi [7] A.H.Mueller and W.-K.Tang, Phys. Lett. {\bf B284} (1992) 123.

\noi [8] H1 Collaboration, S.Aid et al, preprint DESY 96-039 (hep-ex/9603004)
(submitted to Nucl. Phys. {\bf B}).

\noi [9] R.L.Cool et al, Phys. Rev. Lett. {\bf 47} (1981) 701.

\noi [10] K.H.Streng, Phys. Lett. {\bf B166} (1986) 443;
R.Engel, M.A.Braun, C.Pajares and J.Ranft, Santiago preprint
US-FT/18-96, Siegen preprint SI 96-04 (hep-ph/9605227) (submitted to Z. Phys.
{\bf C}).
\newpage

\centerline{\large \bf Figure captions}
\vskip 0.5cm

\noi{\bf Fig. 1.} Diagram corresponding to the absorptive part for central
diffractive events in hadronic (photonic) interactions. Ladders show
pomerons emitted by the projectile and target.\\

\noi{\bf Fig. 2.} The central part of the diagram of Fig. 1. The upper and lower
pomerons are joined to the central one via two Bartels' vertices.

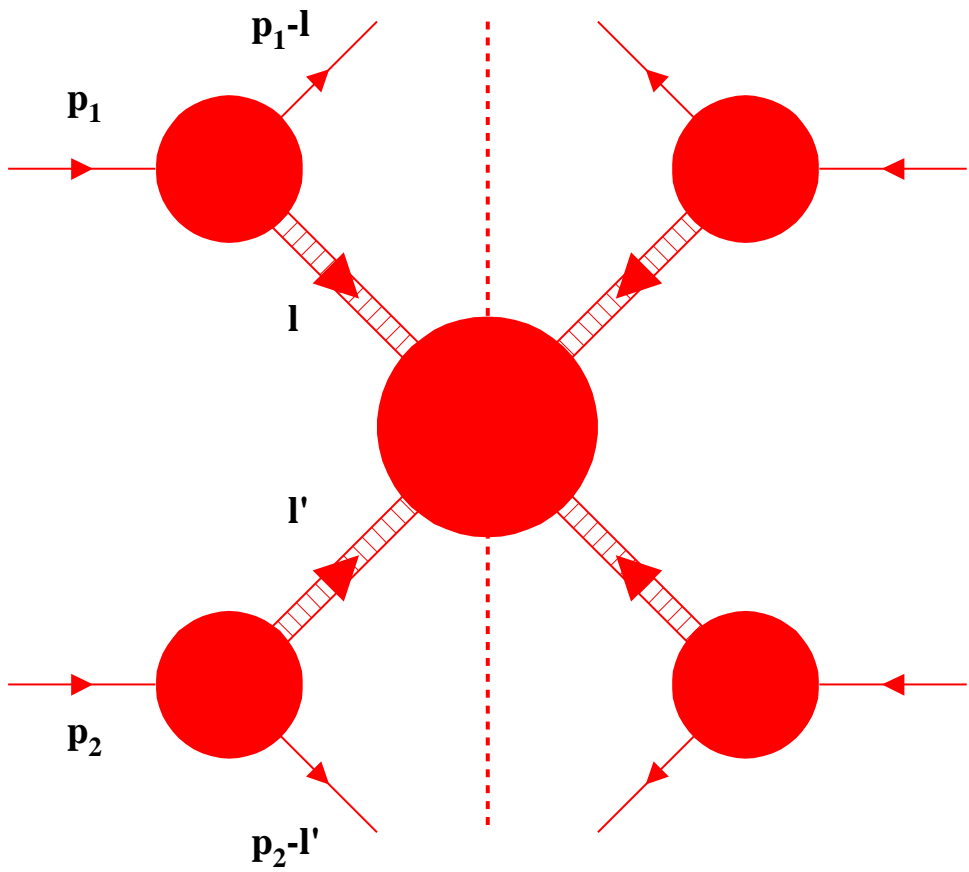
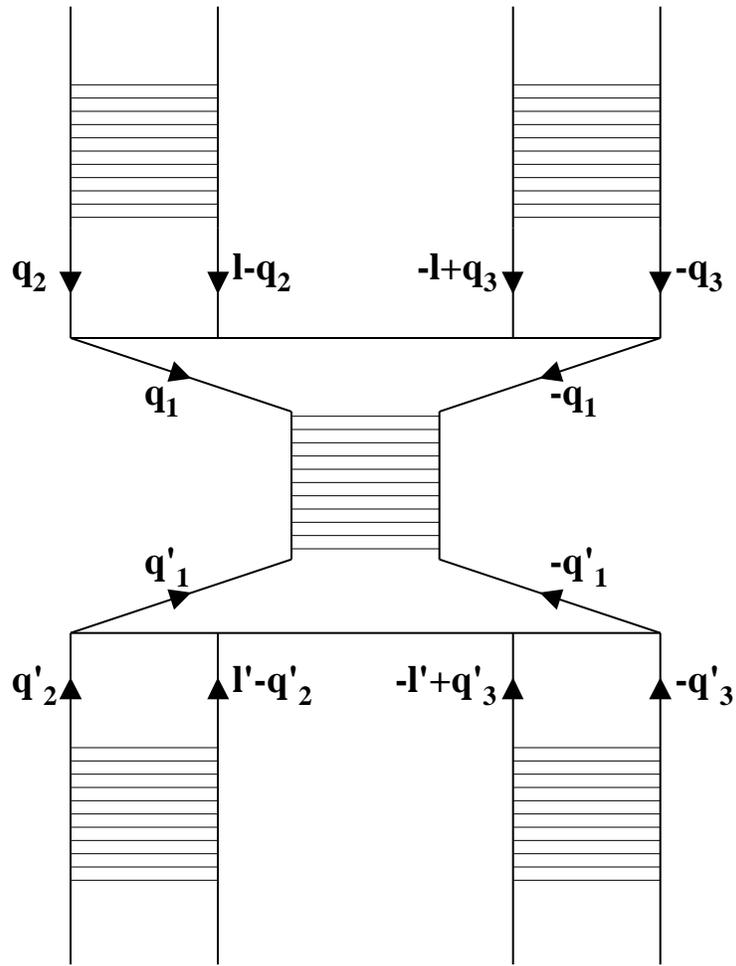
\end{document}